\documentclass[aps,pre,twocolumn,times,superscriptaddress,floatfix]{revtex4}

\usepackage{natbib}

\usepackage{epsfig}

\newlength{\FIGSIZE}

\newlength{\figsize}
\newcommand{\be}{\begin{equation}}
\newcommand{\ee}{\end{equation}}
\newcommand{\ba}{\begin{eqnarray}}
\newcommand{\ea}{\end{eqnarray}}

\newcommand{\la}{\langle}
\newcommand{\ra}{\rangle}

\newcommand{\lr}[1]{\langle #1 \rangle}

\newcommand{\bi}[1]{Fig.~\ref{fig:#1}}
\newcommand{\sect}[1]{section~\ref{sec:#1}}
\newcommand{\Sect}[1]{Section~\ref{sec:#1}}
\newcommand{\eq}[1]{Eq.~(\ref{eq:#1})}

\begin{document}
\title{A comparative study of different integrate \& fire neurons:
  spontaneous activity, dynamical response, and stimulus-induced correlation}
\author{Rafael D. Vilela}
\affiliation{ Max-Planck-Institut f\"ur Physik Komplexer Systeme,
  N\"othnitzer Str.~38 01187 Dresden, Germany}
\affiliation{Centro de Matem\'atica, Computa\c c\~ao e Cogni\c c\~ao, Universidade Federal do ABC, 09210-170, Santo Andr\'e, SP, Brazil}
\author{Benjamin Lindner}
\affiliation{ Max-Planck-Institut f\"ur Physik Komplexer Systeme,
  N\"othnitzer Str.~38 01187 Dresden, Germany}
 \date{\today}

%
%
\setlength{\baselineskip}{1em}
\begin{abstract}
  Stochastic integrate \&  fire (IF) neuron models have found
  widespread applications in computational neuroscience.  Here we
  present results on the white-noise-driven perfect, leaky, and
  quadratic IF models, focusing on the spectral statistics (power
  spectra, cross spectra, and coherence functions) in different
  dynamical regimes (noise-induced and tonic firing regimes with low
  or moderate noise).  We make the models comparable by tuning
  parameters such that the mean value and the coefficient of variation
  of the interspike interval match for all of them. We find that,
  under these conditions, the power spectrum under white-noise
  stimulation is often very similar while the response characteristics,
  described by the cross spectrum between a fraction of the input
  noise and the output spike train, can differ drastically. We also
  investigate how the spike trains of two neurons of the same kind
  (e.g. two leaky IF neurons) correlate if they share a common noise
  input. We show that, depending on the dynamical regime, either two
  quadratic IF models or two leaky IFs are more strongly correlated.
  Our results suggest that, when choosing among simple IF models for
  network simulations, the details of the model have a strong effect
  on correlation and regularity of the output.
\end{abstract}
\maketitle
\section{Introduction}\label{sec:intr}

Neurons communicate information via short-lasting discharges of the electrical
potential across their membrane. The excitability mechanism by which these
spikes are generated relies on the dynamics of voltage-gated ion channels in
the neural membrane and is well understood \cite{Hil01,Koc99}. To study the
dynamics of large neural networks, a detailed description of the single
neuron's dynamics, although in principle possible, is impractical and one must
resort to simpler models of neural spike generation governed by only one or
two dynamical variables per neuron \cite{GerKis02}. In particular in
stochastic versions, that take into account the large variability of neural
spiking, these models can be also helpful to study basic aspects of signal
transmission by single neurons.

One class of commonly used simplified models comprises integrate~\&~fire (IF)
neurons with white noise current. In IF models a spike is generated if the
voltage reaches a firing threshold (inducing also a reset of the voltage); the
voltage obeys the one-dimensional dynamics
\begin{equation}
\label{eq:if}
\dot{v}=f(v)+s(t)+ \mu + \sqrt{2D} \xi(t)
\end{equation}  
where $s(t)$ is a time-dependent stimulus while $\mu$ and $D$ are the
mean and the noise intensity of the input current ($\xi(t)$ is white
Gaussian noise).  Variants of the model differ by the function $f(v)$.
A fine-tuned choice of $f(v)$ may permit a rather accurate prediction
of both experimental subthreshold voltage fluctuations and spike
statistics under noisy stimulation {\em in vitro} and {\em in vivo}
(see \cite{BadLef08,JolKob08,JolSch08,RauLaC03,LanSan06} for some recent
convincing examples).  Simple choices like a constant, linear, or
quadratic function leading to the perfect (PIF), leaky (LIF), or
quadratic (QIF) model, respectively, allow for an analytical
calculation of one or the other spike statistics and may be also
numerically simpler to simulate in large-scale networks.  Models like
\eq{if} have been used in analytical studies of (i) conditions for
asynchronous or oscillatory activity in recurrent networks
\cite{AbbVre93,AmiBru97b,Bru00,HanMat01}; (ii) the transmission of
rapid signals \cite{BruCha01,LinLSG01,FouHan03,NauGei05}; (iii) the
variability of spontaneous activity
\cite{RicSac79,GutErm98,LinLSG02,LinLon03}; (iv) noise-induced
resonances in the spontaneous activity \cite{PakTan01,LinLSG02}, and
in the response to external stimuli \cite{Ste96,LinLSG01}; and (v)
oscillations in recurrent networks induced by spatially correlated
stimuli \cite{DoiCha03,DoiLin04,LinDoi05}, to name but a few.

Most of the phenomena studied depend strongly on the choice of
$f(v)$. As an example, consider the effect of coherence resonance,
which refers to the existence of an optimal noise intensity $D$ that
maximizes the regularity of the spike train, seen, for instance, as a
minimum of the coefficient of variation (CV) vs noise intensity: only
the leaky \cite{PakTan01,LinLSG02} but not the perfect or quadratic IF
models \cite{LinLon03} show such a minimum. It has, furthermore, been
shown that LIF and QIF differ strongly in their response to fast
(high-frequency) periodic signals \cite{FouHan03,NauGei05}. Despite
these discussions, however, a systematic comparison among the commonly
used IF models is still missing. In this paper, we want to fill this
gap.

If one wishes to compare different IF models, the first question is how
the input parameters $\mu$ and $D$ should be chosen in the respective
model. Already the most basic firing statistics of a certain IF model, the
firing rate (quantifying the spike train's intensity) and the interspike
interval's coefficient of variation (characterizing the variability of
the spiking) depend strongly and in a model-specific way on $\mu$ and $D$
\cite{Ric77,RicSac79,LinLSG02,LinLon03,Bur06}.  The authors have recently
shown that this basic firing statistics, i.e. rate and CV, determine uniquely
the input parameters $\mu$ and $D$ for the three most common IF models
mentioned above (perfect, leaky, and quadratic IF neurons). This offers a
natural way of unambiguously defining firing regimes for these models.
Moreover, setting the firing regime by means of prescribing rate and CV allows
for a fair comparison of the higher-order statistics of different IF
models. In this way, we can, for example, consider an LIF neuron and a QIF
neuron that both show a moderate firing rate (e.g. 10Hz) and medium
variability (say, a CV of about 0.5) and compare how these two neurons differ
in their spontaneous and driven activity. This approach of assuming the same
basic firing statistics and comparing higher-order statistics is complementary
to a previous set-up which was entirely based on the firing rate dependence
on input current \cite{FouHan03}.

What is the most important output statistics of noisy IF neurons once the
firing rate and CV are fixed?  In most of the above analytical approaches, two
single-neuron characteristics appear to be essential: (i) the spike train
power spectrum of spontaneous activity and (ii) the response to weak stimuli
(e.g. to a weak periodic signal $s(t)=\varepsilon \cos(\omega t)$). In a more
recent theory of recurrent networks \cite{LinDoi05}, the knowledge of a third
simple property is needed that goes beyond the properties of a single neuron:
the degree of correlations that can be induced in two uncoupled neurons that
share some common noisy stimulus. This latter property has attracted attention
of its own and has been recently studied experimentally (see \cite{RocDoi07}
and references therein) and theoretically \cite{SalSej01,LinDoi05,SheJos08} in
particular in the limit of a weak input correlation.

In the present paper, we study the spontaneous power spectrum, the linear
response function (susceptibility), and the two-neuron correlations induced by
a common stimulus for the perfect, leaky, and quadratic IF models and a
variety of firing regimes.  In \sect{models}, we introduce the three IF models
studied and define the firing regimes. In \sect{spont}, we present results on
spontaneous activity of single neurons. We show that typically IF neurons 
present similar power spectra when they are in the same firing
regime. In \sect{resp}, we study the dynamical response of single neurons. We
recover characteristic differences between the susceptibility of the LIF and
QIF discussed previously (see, e.g. \cite{FouHan03}), as well as between LIF
and PIF \cite{SteFre72}, and show in addition that the spectral coherence
between spike train and external signal is basically low-pass for all three
models. \Sect{2n} is devoted to the study of two neurons driven in
part by common noise.  In this case, linear response theory leads to a good
approximation for the cross-spectra between the two output spike trains when
the common noise makes up only a small fraction of the total noise.  Coherence
functions of the two output spike trains are again low-pass and resemble
qualitatively the input-output coherence functions discussed before. Finally,
we discuss the correlation coefficient of the spike count for the LIF and QIF
models for a weak common noise and show analytically in the appendix that this
correlation coefficient is equal to the input correlation for the PIF model.
We summarize our results and draw some conclusions in \sect{conc}.


\section{Models and firing regimes}\label{sec:models}

\subsection{Integrate-and-fire neuron models}
In this paper we consider IF neurons  subjected to stochastic
voltage-independent input current, i.e., additive white Gaussian noise which 
can be justified in the so-called diffusion approximation
\cite{Hol76,Ric77,Tuc89}. We will consider exclusively models driven by white
noise, setting the term $s(t)$ in \eq{if} to zero; a fraction of the input noise will
later be regarded as a stimulus or as common noise.

For a leaky integrate-and-fire neuron, the current balance equation  reads 
\ba
\label{eq:lif_specific}
&&C_m \dot{V}=-g_L(V-V_L)+ \bar{\mu} + \sqrt{2\bar{D}} \xi(t),\\[.5em]
&& \mbox{if}\;\;\;\; V(t)=V_{th}\;\;\; \mbox{then spike at $t_i=t$ \& } \;\; V \to V_{r} \nonumber
\ea
where $C_m$ is the capacitance of the cell membrane, $g_L$ and $V_L$ are leak conductance and leak reversal potential, respectively, and $\bar{\mu}$ and $\bar{D}$ are the mean and the intensity of the white Gaussian input noise current. The second line describes the fire-and-reset rule upon reaching the threshold $V_{th}$. 

With the simple transformation $v= (V-V_L)/(V_{th}-V_{r})$  and the new parameters $\tau_m=C_m/g_L$
(membrane time constant), $\mu=\bar{\mu}/g_L$, and $\hat{D}=\bar{D}/g_L^2$, this reads
\ba
\label{eq:lif_standard}
&&\tau_m \dot{v}=-v+ \mu + \sqrt{2\hat{D}} \xi(t),\\[.5em]
&& \mbox{if}\;\;\;\; v(t)=v_{th}\;\;\; \mbox{then spike at $t_i=t$ \& } \;\; v \to v_{r} \nonumber
\ea
where the threshold and reset are now at $v_{th}=1$ and $v_r=0$. When measuring time in multiples of the membrane time constant, i.e. $\hat{t}=t/\tau_m$, this model is equivalent to \eq{if} with a rescaled noise intensity $D=\hat{D}/\tau_m$ and with $f(v)=-v$. Note that  $\mu+\sqrt{2D}\xi(t)$ in this rescaled model has not the physical dimensions of an electric current anymore and that is why  we will refer to it here   with the more general 
term 'input'.
 
If the leak term $g_L(V-V_L)$  is small compared to the mean input current, we may be justified to neglect it. All previous transformations  can be repeated  (including the division by the leak conductance $g_L$) and thus we end up with 
\ba
\label{eq:pif_standard}
&&\tau_m \dot{v}=\mu + \sqrt{2\hat{D}} \xi(t),\\[.5em]
&& \mbox{if}\;\;\;\; v(t)=v_{th}\;\;\; \mbox{then spike at $t_i=t$ \& } \;\; v \to v_{r} \nonumber
\ea
which corresponds after rescaling of time again to \eq{if} but this time with $f(v)\equiv 0$. This is the perfect integrate-and-fire (PIF) model with white noise  (also known as random-walk model of neural firing) \cite{GerMan64,Hol76}.

If the noise-free neuron is close to a dynamical bifurcation, specifically, close to  a saddle-node bifurcation from quiescence to tonic firing, another form of the integrate-and-fire neuron contains a quadratic nonlinearity \cite{Erm96,GutErm98,LatRic00,HanMat01,LinLon03,Izh07}
\ba
\label{eq:qif_specific}
&&C_m \dot{V}=a(V-V_0)^2+ \bar{\mu} + \sqrt{2\bar{D}} \xi(t),\\[.5em]
&& \mbox{if}\;\;\;\; V(t)=\infty \;\;\; \mbox{then spike at $t_i=t$ \& } \;\; V \to -\infty \nonumber
\ea
In this case one chooses threshold and  reset at infinity because the slow dynamics in $V$ makes the exact
(large but finite) values of $V_r$ and $V_{th}$ irrelevant. Note also that $V$ in this case can be but has not to be  a voltage --- in general, it is the  variable of the center manifold \cite{Izh07}; correspondingly, the factor $C_m$ on the left-hand-side can be taken as a convention.  For infinite reset and threshold values, this dynamics can be brought into a simplified standard form by choosing a new variable $v=a(V-V_0)/g_L$ and new parameters $\mu=a\bar{\mu}/g_L^2$ and $\hat{D}=\bar{D} a^2/g_L^4$:
\ba
\label{eq:qif_standard}
&&\tau_m \dot{v}= v^2 +\mu + \sqrt{2\hat{D}} \xi(t),\\[.5em]
&& \mbox{if}\;\;\;\; v(t)=\infty \;\;\; \mbox{then spike at $t_i=t$ \& } \;\; v \to \infty \nonumber
\ea
which corresponds  in rescaled time $\hat{t}=t/\tau_m$ and noise intensity $D=\hat{D}/\tau_m$ to \eq{if} with $f(v)=v^2$. In simulations of this standard form of this quadratic integrate-and-fire neuron (QIF), one uses large but finite  threshold and reset such that -- by further increasing their values -- the results (ISI statistics, spike train power spectra, etc.) do not change anymore within the desired accuracy (say, curves do not change in line thickness). For the effect of finite values of reset and threshold values on the ISI statistics, see \cite{LinLon03}.   
  
Note that both in the PIF and the QIF the introduction of the membrane time scale is  arbitrary  --- we could equally well compare to PIF and QIF models in which $\tau_m$ would be replaced by a multiple or a fraction of this time (changing then also the input parameters, of course). The choice of $\tau_m$ has been made previously
for the PIF \cite{FouBru02} and we follow here this convention also for the QIF.

Our approach of comparing different IF models here is complementary to
others in which the input current is assumed to be known and the
parameters of the specific models (e.g. \eq{lif_specific} and
\eq{qif_specific}) are fitted to yield a given ISI statistics. Here we
start with the standard models \eq{lif_standard}, \eq{pif_standard},
and \eq{qif_standard} and ask for the input parameters that yield a
given rate (in units of the inverse membrane time constant) and a
given CV. Although this may seem to be unusual in an experimental
situation where one has control over the input current, it appears to
be a reasonable approach {\em in vivo} where the effective input
current and its noise intensity is set by the synaptic background and
thus unknown.

\subsection{Firing regimes}\label{subsec:regimes}
\begin{figure}[h]
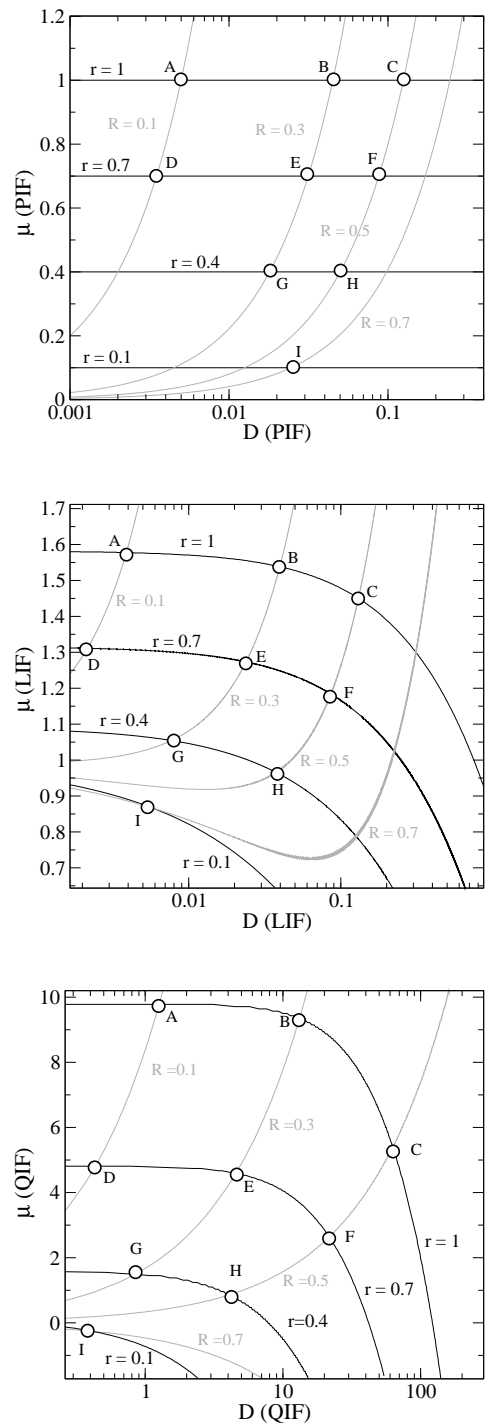
 
\includegraphics[width=0.35\textwidth]{pif.eps}\\ \vspace{7mm}
\includegraphics[width=0.35\textwidth]{lif.eps}\\ \vspace{7mm}
\includegraphics[width=0.35\textwidth]{qif.eps}
\caption{
\label{fig:contours} 
Contour lines for rate (black) and CV (gray) in parameter space for the different models.
Regimes A-I are defined by intersections of these contour lines.
}
\end{figure}

In order to make different IF models comparable, we must first specify the
correspondence between their parameters. For instance, in a comparison between
LIF and QIF, we should first decide which pair $(D,\mu)$ for the first model
corresponds to which pair $(D,\mu)$ for the second.  Here we do this by
defining different firing regimes in terms of fixed rate and coefficient
of variation of the spike trains.  In order words, $D$ and $\mu$ in
different models are chosen as to yield a given firing rate 
\be
r=\frac{1}{\langle T \rangle}
\label{eq:rate}
\ee
and a given coefficient of variation
\be
R=\frac{\sqrt{\langle T^2 \rangle - \langle T \rangle^2}}{\langle T \rangle},
\label{eq:cv}
\ee 
where $T$ is the interspike interval.  Throughout this paper
  $\la \cdot \ra$ denotes averaging over realizations of the
  stochastic process.  Note that, since time is measured in units of
the membrane time constant, all the rates are in units of the inverse
of this constant. For instance, for a membrane time constant of 10ms,
a nondimensional rate of 1 corresponds to 100Hz.

The pair of parameters $(D,\mu)$ for a given model that yields a certain
regime is therefore determined by the intersection between one countour line
for the rate and one contour line for the CV. However, it is not clear {\it a
priori} whether at most one such intersection exists. This problem was
recently addressed by us \cite{VilLin09}. We showed that, given fixed rate and
CV, there can be at most one associated pair $(D,\mu)$ for the three IF models
studied in this paper.

\bi{contours} displays some contour lines for the rate and CV for the three
models considered here. 
There are different ways to determine these contour lines. They can be 
obtained analytically (see \cite{VilLin09}). Here we have determined them
numerically, as explained in \sect{spont_res}.
In \bi{contours} we also define 9 specific regimes, labeled A-I, 
which we study in some detail here. The corresponding values for the
rate and CV in these regimes are:

\begin{center}
  \begin{tabular}{ | c || c | c | c | c | c |  c | c | c | c |}
    \hline
    Regime & A & B & C & D & E & F & G & H & I \\ \hline 
    rate & 1 & 1 &1 & 0.7 & 0.7 & 0.7 & 0.4 & 0.4 & 0.1 \\ \hline
    CV & 0.1 & 0.3 & 0.5 & 0.1 & 0.3 & 0.5 & 0.3 & 0.5 & 0.7 \\ \hline
  \end{tabular}
\end{center}



\section{Single neurons: spontaneous activity} \label{sec:spont}

\subsection{Measures}

The output spike train $y$ can be modelled as a sum of delta peaks at the time
instants when the voltage described by \eq{if} reaches the threshold value:
\be
y (t)=\displaystyle\sum_j \delta(t-t_{j}),
\ee
where $t_{j}$ is the instant when the $j$-th spike occurs.

The spontaneous activity of the IF neurons studied here corresponds to a
renewal point process. Each interspike interval is an independent random
variable. In processes of this type, all the information on the statistics is
contained in the probability density of the ISI.

In this paper, we will quantify the neuron's correlation statistics by means of power and cross-spectra. We start by defining the Fourier transform of the zero-average spike train as:
\begin{equation}
  \tilde{y} (f)=\int_0^T dt' e^{2\pi if t'} (y (t')- \langle y (t') \rangle ).
\end{equation}
The power spectrum of the spike train will be the quantity used in this
paper to characterize the spontaneous activity of the IF neurons. It is given
by:
\begin{equation}
S_{y} (f) = \lim_{T\rightarrow \infty} \frac{1}{T} \langle \tilde{y}
\tilde{y}^* \rangle,  \label{eq:sy}
\end{equation}
where $T$ is the realization time window.  For renewal point processes, the
relation between the power spectrum of the spike train and the Fourier
transform of the probability density of the ISI, $F (f)$, is given by
\cite{CoxLew66}:
\be
S_y (f) = \frac{1}{\lr{T}} \frac{1-|F(f)|^2 }{|1-F(f)|^2 }.
\label{eq:spec_pdf}
\ee
We note that analytical expressions for $F (f)$ are known in the cases
of PIF and LIF (equivalently, often the Laplace transform is stated
that yields the Fourier transform for a negative imaginary argument).
In this work, we will only use that for the PIF, which is given by
\cite{SugMoo70,SteFre72}:
\be
F^{(\mbox{PIF})}(f) = \mbox{exp}\left[(v_{th} -v_{r})\left(\frac{\mu}{2D} -\sqrt{\frac{\mu^2}{4D^2}-\frac{2if\pi}{D}}\right)  \right].
\label{ps_pif}
\ee

\subsection{Results}\label{sec:spont_res}

\begin{figure}[h]
\includegraphics[width=0.45\textwidth]{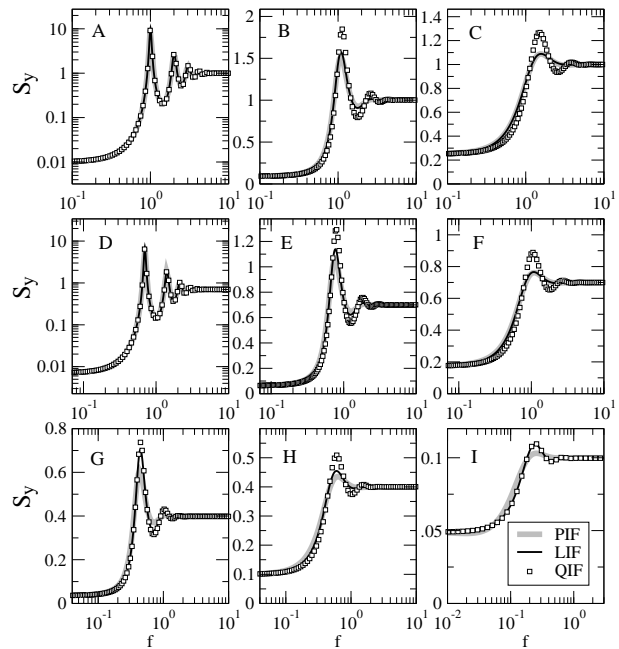}
\caption{
\label{fig:spec9} 
Power spectra of the spike trains for the three models in the 
nine firing regimes defined in Fig. 1.
}
\end{figure}
\begin{figure}[h]
\includegraphics[width=0.45\textwidth]{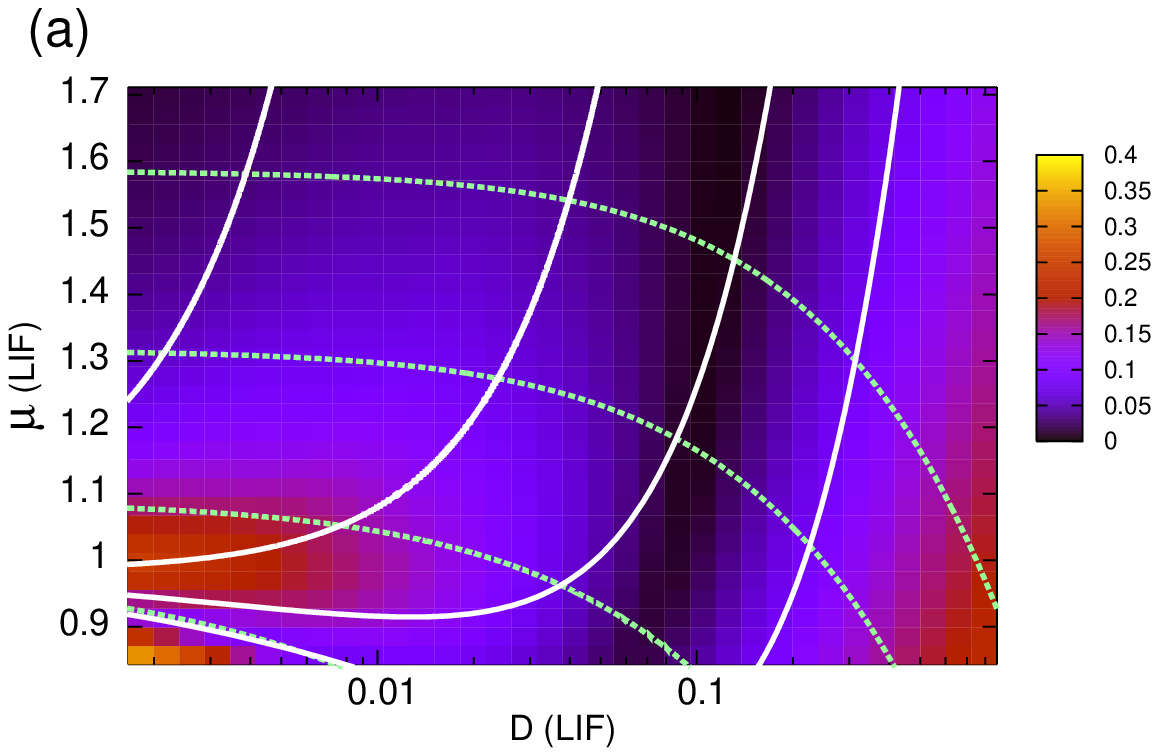}
\includegraphics[width=0.45\textwidth]{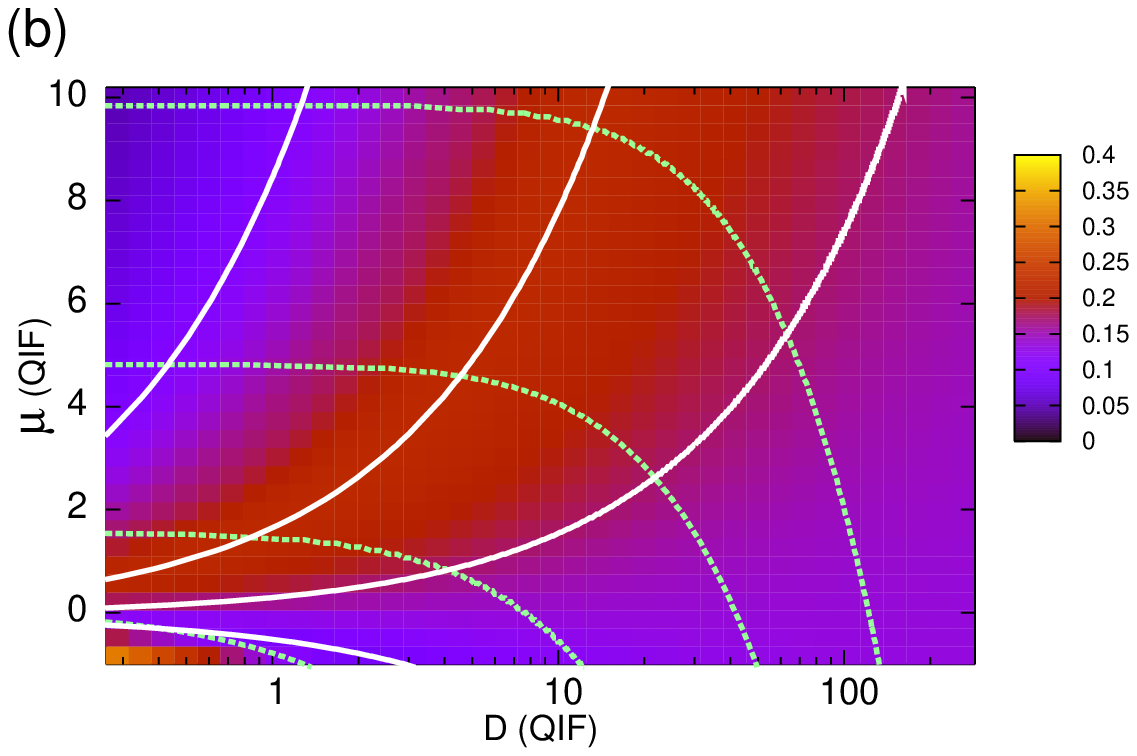}
\includegraphics[width=0.45\textwidth]{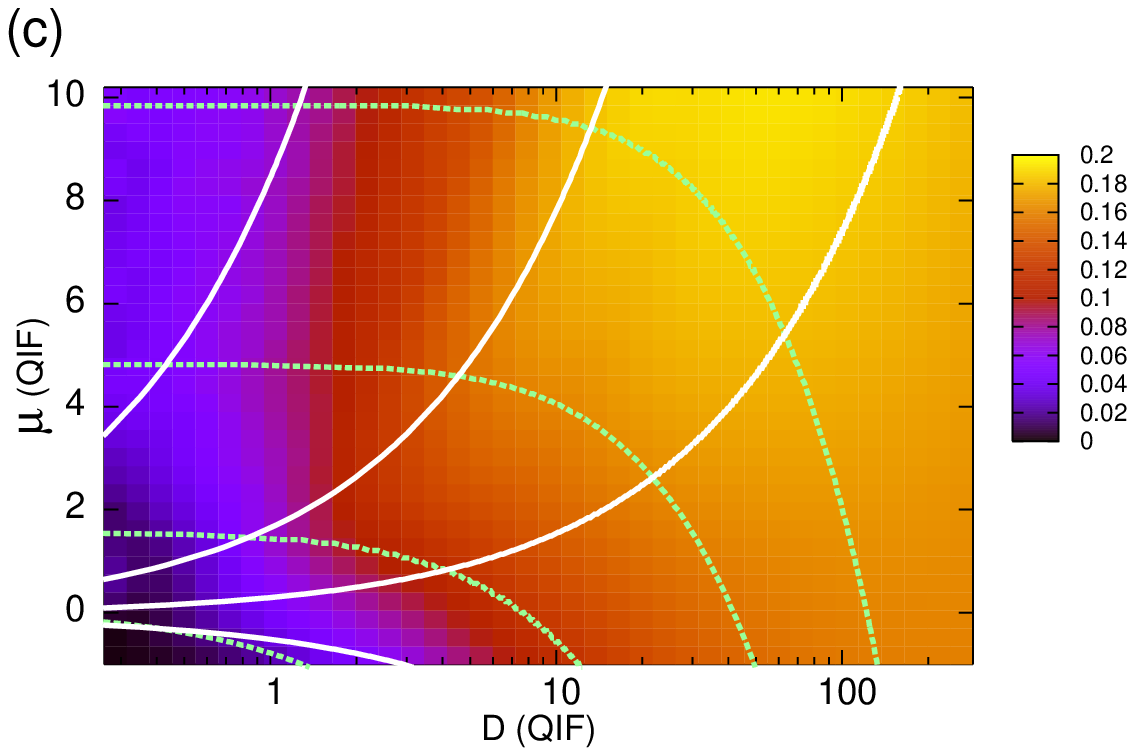}
\caption{
\label{fig:pscomp} 
(Color online) Maximal (over all frequencies) relative difference of power spectra $\Delta
S_{y}^{\small{\mbox{PIF,LIF}}}$ (a), $\Delta S_{y}^{\small{\mbox{PIF,QIF}}}$ (b), and
$\Delta S_{y}^{\small{\mbox{LIF,QIF}}}$ (c).  The contour lines for rate and CV are
the same as those depicted in Fig. 1.  }
\end{figure}

In \bi{spec9}, we show the power spectra for the three models in
regimes A-I.  These spectra were obtained analytically via Eqs.
(\ref{eq:spec_pdf}) and (\ref{ps_pif}) for the PIF and numerically for
LIF and QIF using the algorithm recently introduced by Richardson
\cite{Ric08}.  We observe that the power spectra of different models
in the same regime are in general very similar. In Regimes A and D,
which are characterized by low variability (CV equal to 0.1), the power
spectra virtually coincide. In the other regimes, the power spectra
coincide in the limits of low and large frequencies, and deviate to
some extent in the intermediate-frequency range.

The coincidence of the power spectra for different models in the same regime
in the low and large frequency limits is not surprising.
In fact, for renewal point processes one can show \cite{CoxLew66} that:
\be
\lim_{f \to 0}S_{y}(f)=rR^2 
\label{eq:lim_f_0}
\ee
and
\be
\lim_{f \to \infty}S_{y}(f)=r.
\label{eq:lim_f_infty} 
\ee
Since each regime is defined by fixing the rate and CV, we conclude from
\eq{lim_f_0} and \eq{lim_f_infty} that the power spectra for different models
should indeed coincide in these limits.  In fact, we have used these relations
and the above mentioned algorithm for the numerical determination of the power
spectrum \cite{Ric08} to obtain the contour lines displayed in \bi{contours}.
 
To quantify the differences between the power spectra of different models, we
define the maximal relative difference $\Delta S_{y}^{j,k}$ between the
power spectra of models $j$ and $k$ over all frequencies as:
\be
\Delta S_{y}^{j,k}=\mbox{max}_f \left(\frac{|S_{y}^{(j)}-S_{y}^{(k)}|}{(S_{y}^{(j)}+S_{y}^{(k)})/2}\right).
\ee
In \bi{pscomp} we plot $\Delta S_{y}^{\small{\mbox{PIF,LIF}}}$,
$\Delta S_{y}^{\small{\mbox{PIF,QIF}}}$, and $\Delta
S_{y}^{\small{\mbox{LIF,QIF}}}$.  The first observation we make is
that the power spectra of the PIF matches those of the LIF and QIF in
the parameter regions where the PIF is a good model, i.e., for large
$\mu$ and small $D$ [cf. \bi{pscomp}(a) and (b)]. Second, when
comparing the LIF with the QIF (\bi{pscomp}(c)), we conclude that the
power spectra of these models are practically coincident if the noise
intensity is small and their relative difference increases with $D$,
displaying moderate differences for very large noise intensity.  There
is a nontrivial dependence on $\mu$ as well, but the dependence on $D$
is the dominant one. Remarkably, this rule of thumb whereby the power
spectra differences between LIF and QIF increase with the noise
intensity is valid for both tonic ($\mu_{\mbox{QIF}} >0$) and
noise-induced ($\mu_{\mbox{QIF}} < 0$) firing regimes.


\section{Single neurons: dynamical response} \label{sec:resp}

\subsection{Measures}
In this section, we are interested in the response of single neurons to a
small stimulus.  This can be accomplished in several ways, e.g., by adding a
small term with sinusoidal time dependence to \eq{if}. Alternatively, and this
is the procedure adopted here, one can regard a fraction of the noise term in
\eq{if} as the external stimulus. This choice will allow for a straightforward
connection between the single neuron's response presented in this section and
two-neuron correlations under common noise discussed in \sect{2n}. We thus
rewrite \eq{if} as:
\begin{equation}
\dot{v}=f(v)+\mu+  \sqrt{2(1-c)D}  \xi_i (t)+  \sqrt{2cD}\xi_c (t),
\label{eq:if2}
\end{equation}
where the noise terms $\xi_i (t)$ and $\xi_c (t)$ are white Gaussian and
described by:
$$
\langle \xi_i (t)\rangle=\langle \xi_c (t)\rangle=\langle \xi_i (t)  \xi_c
(t')\rangle=0,
$$
\begin{equation}
\langle \xi_c (t) \xi_c (t')\rangle=\langle \xi_i (t) \xi_i
(t')\rangle=\delta(t-t'),
\label{eq:noise1}
\end{equation}
and $c$ (playing the role of a relative signal amplitude) is a number between
0 and 1.  When addressing the single neuron's response, we read \eq{if2} as
describing a certain neuron subjected to  intrinsic noise $\sqrt{2(1-c)D}
\xi_i (t)$ and driven by an external (noisy) stimulus:
\be
s(t)= \sqrt{2cD}\xi_c (t).
\label{eq:stim}
\ee

To characterize the neuron's response to the stimulus, we use the cross-spectrum between  the spike train $y$ and the
stimulus
$s(t)$,
\begin{equation}
S_{y s} (f) = \lim_{T\rightarrow \infty} \frac{1}{T} \langle \tilde{y}
\tilde{s}^* \rangle,      \label{syxi}
\end{equation}
and the coherence function with respect to $s$,
\begin{equation}
\gamma^2 (f) = \frac{|S_{y s}|^2}{S_{y}S_{s}},  \label{coher}
\end{equation}
where $S_{s}=2cD$ is the power spectrum of $s$.  One should note that the
coherence function is restricted to the interval $0< \gamma^2 (f) <1$.

\subsection{Single neuron's response}

The cross-spectra (\ref{syxi}) can be calculated, for small $c$, from
linear
response theory.  The idea is to consider the term $\sqrt{2cD}\xi_c$ as a small
perturbation of the term $\mu$ in the stochastic system
\begin{equation}
\dot{v}=f(v)+\mu+\sqrt{2(1-c)D}\xi_i (t). \label{unpert}
\end{equation}
This does not seem feasible at first sight, since $\xi_c $ has infinitely
large variance.  To show that linear response can be applied in this case,
we
formally consider $\xi_c$ as a band pass white noise with flat
spectrum of
height $2cD$ and cutoff frequency $f_{max}$.  Its variance is then equal
to
$2cDf_{max}$.  This variance can be kept small even in the limit of
$f_{max}
\rightarrow \infty$ (white noise) if we impose that $c$ decreases
sufficiently
fast, i.e., $c\ll \mu/2Df_{max}$.  Therefore $\xi_c $ can be indeed
considered
a small perturbation.  Linear response theory \cite{Ris84} then leads to
the
following approximation:
\begin{equation}
\langle \tilde{y} (f)\rangle=\chi_{D,\mu} (f)
\sqrt{2cD} \tilde{\xi}_{c}(f), \label{eq:lin_resp}
\end{equation}
where $\chi_{D,\mu}$ is the {\it susceptibility}
of the system which can be estimated from the cross spectrum between
input signal and spike train via the well-known relation 
\begin{equation}
\chi_{D,\mu}=\frac{S_{y s} (f)}{2cD}= \frac{\lim_{T\rightarrow \infty} \langle \langle \tilde{y}
\sqrt{2cD}\tilde{\xi}_{c}^* \rangle_{\xi_i} \rangle_{\xi_c}}{2cD},      \label{eq:syxi2}
\end{equation}

Closed analytical forms for $\chi$ exist for the PIF \cite{FouBru02}
and LIF \cite{LinLSG01,BruCha01}, and are given respectively by:
\begin{equation}
\chi_{PIF}= \frac{\mu^2}{v_{th} -v_r  } \frac{1-\sqrt{1-8\pi ifD/\mu^2}}{4\pi i fD}
\label{eq:susc_pif}
\end{equation}
and
\begin{equation}
\chi_{LIF}=\frac{r2\pi if/\sqrt{D}}{2\pi if-1}\frac{\mathcal{D}_{2\pi if
    -1}\left(\frac{\mu-v_{th}}{\sqrt{D}}\right)
-e^\delta \mathcal{D}_{2\pi if -1}\left(\frac{\mu-v_r }{\sqrt{D}}\right) }
{\mathcal{D}_{2\pi if}\left(\frac{\mu-v_{th}}{\sqrt{D}}\right)
-e^\delta \mathcal{D}_{2\pi if}\left(\frac{\mu-v_r }{\sqrt{D}}\right)},\label{eq:susc_lif}
\end{equation}
where the rate $r$ for the LIF is given by
\be
r(\mu,D)=\left(\sqrt{\pi}\int_{(\mu-v_{th})/\sqrt{2D}}^{(\mu-v_r )/\sqrt{2D}}dz e^{z^2}\mbox{erfc}(z)\right)^{-1},
\ee
the abbreviation $\delta$ reads
\be
\delta=\frac{v_{R}^2-v_{T}^2+2\mu(v_{th} -v_r  )}{4D},
\ee
and $\mathcal{D}_a (z)$ is the parabolic cylinder function \cite{AbrSte70}.
For the QIF, $\chi$ can be obtained numerically from the Fokker-Planck
equation \cite{Ric07}.
\subsection{Results}
\begin{figure}[h]
\includegraphics[width=0.45\textwidth]{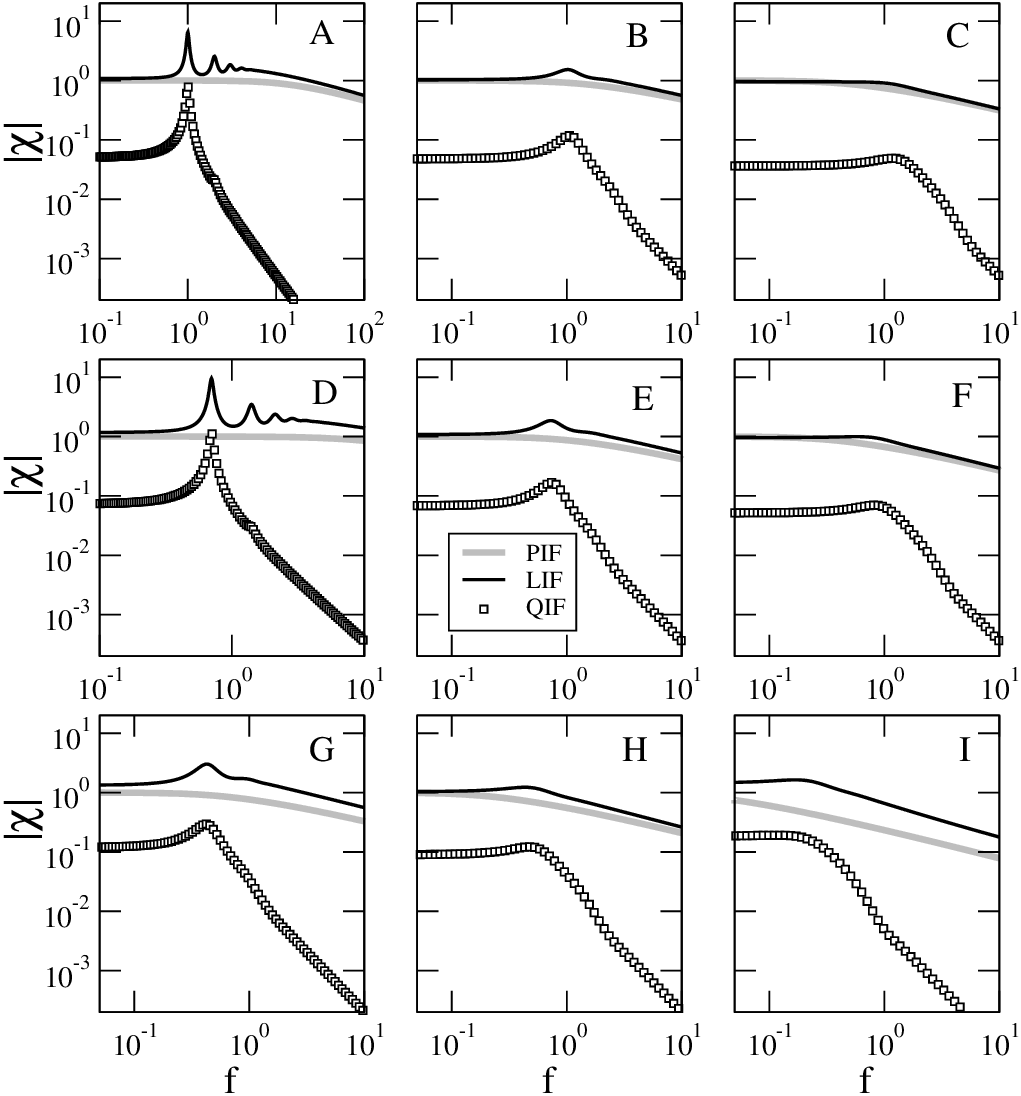}
\caption{
\label{fig:gain} 
Gain ($|\chi|$) as a function of frequency for the three models 
in the Regimes A-I.
}
%
\includegraphics[width=0.45\textwidth]{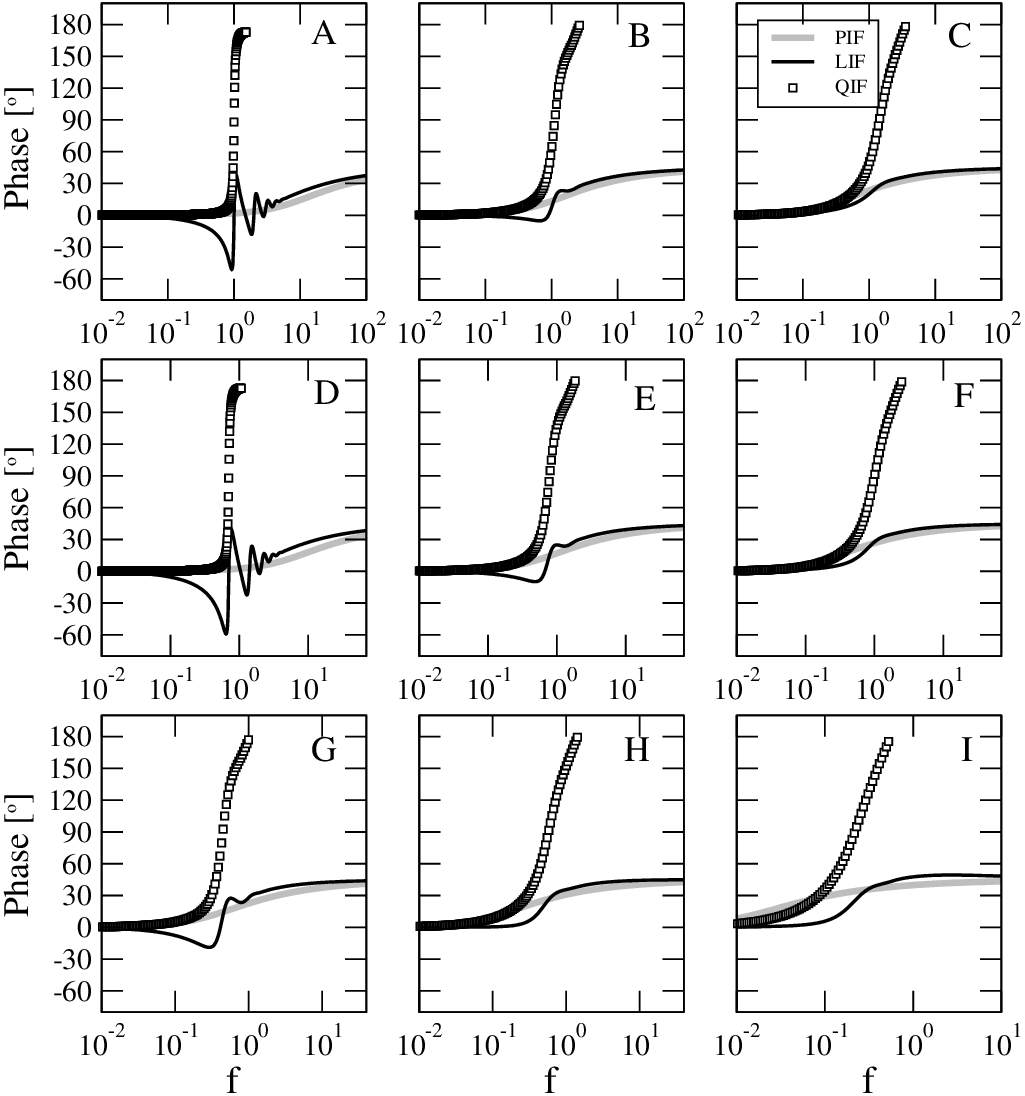}
\caption{
\label{fig:phase} 
Phase lag of the linear response (i.e. $-\arg(|\chi|)$) as a function of frequency for the three models 
in the Regimes A-I. Due to numerical constraints, the phase of the QIF model 
is not shown in the large frequency range, where it asymptotes to $180^{\mbox{o}}$. 
}
\end{figure}

Since \eq{syxi2} is valid for arbitrary $c$, the cross-spectrum $S_{y s}$ is
fully characterized by the susceptibility $\chi_{D,\mu}$.  We have studied the
susceptibility for the three models in regimes A-I.  The susceptibility for
the PIF was determined using \eq{susc_pif}, while the susceptibility for the
LIF and QIF was determined by integrating the Fokker-Planck equation with the
algorithm presented in \cite{Ric07, Ric08}.  It turns out that this numerical
integration is faster than the evaluation of \eq{susc_lif} using standard
softwares.

In \bi{gain}, we show the gain $|\chi|$ as a function of the frequency for the
three models in Regimes A-I. The gain is typically larger for the LIF and is
in all regimes at least one order of magnitude smaller for the QIF. In regimes
A and D, where the firing is most regular, the gain displays peaks for the LIF
(close to the firing frequency and its higher harmonics)
and QIF (only close to its firing frequency) but not for the PIF.  As observed in \cite{FouHan03}, in the large
frequency limit the gain decays as a power law with exponent 0.5 for the LIF
and 2 for the QIF. From \eq{susc_pif}, we see that the exponent for the PIF is
also 0.5.

In \bi{phase}, we display the phase $\phi$ for the different models.
It is defined such that $\chi=|\chi|e^{i\phi}$. For the PIF, it is in
all regimes close to zero for small frequencies and increases
monotonically. Its saturation value, attained in the limit $f\to
\infty$, is $\phi=45^o$.  Except for the limits of small and large
frequencies, the behavior of the LIF can be markedly different. In
particular in Regimes A and D, the phase oscillates around zero in a
certain range enclosing the eigenfrequency.  It first becomes
negative. Close to the eigenfrequency it changes signal, and repeats
this oscillatory behavior a few times before approaching its
asymptotic value of $45^o$. Finally, the behavior of the phase for the
QIF is similar to the one of the LIF, except that the asymptotic value
at large frequencies is remarkably larger - equal to $180^o$. The
asymptotic behaviors for the LIF and QIF were also discussed in
\cite{FouHan03}.
\begin{figure}[h]
\includegraphics[width=0.45\textwidth]{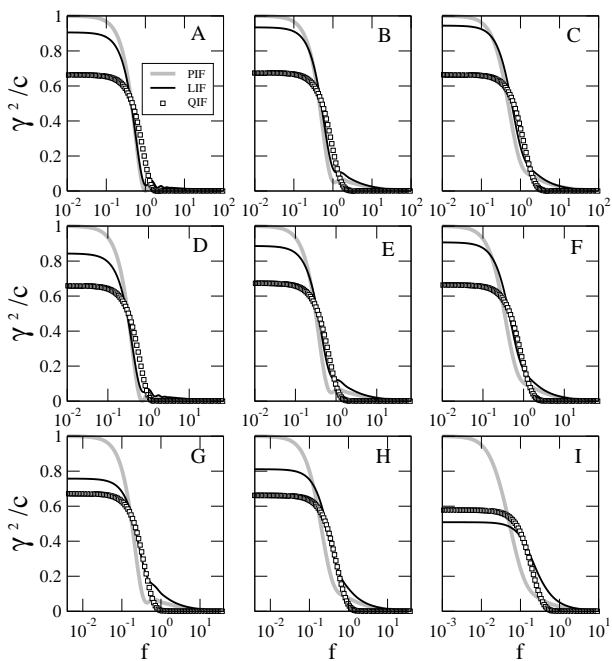}
\caption{
\label{fig:coh} 
Coherence function $\gamma^2$ for the three models 
in the Regimes A-I. Note that, in Regime I, the coherence of the QIF
is larger than that of LIF for small frequencies.
}
\end{figure}

We now turn to the coherence function $\gamma^2$  that we show in multiples  of $c$ for the different models in  \bi{coh}.  Although the gain of the three models differed by more than one order of magnitude and showed different resonances, their coherence functions 
are rather similar and display a  low-pass behavior.  Therefore PIF, LIF,
and QIF transmit most information in a low-frequency band of the stimulus.
Going to the limit of vanishing frequencies, the PIF will transmit most information:
as can be explicitely shown \cite{SteFre72},  $\gamma^2 $ for the PIF approaches the
maximum value $c$ in the limit of zero frequency, a feature not shared by
neither LIF nor QIF.  Furthermore, the coherence function at low frequencies  can be larger for
the LIF as compared to the QIF (regimes A-H) and, conversely, larger for the
QIF as compared to the LIF (regime I).  As we will argue in the next section,
this feature also affects which of these models will display larger two-neuron
correlations under common noise stimulus.


\section{Two-neuron correlations under common stimulus} \label{sec:2n}

\subsection{Measures} \label{sec:2nA}

We now study two neurons of the same model subjected to individual noise and
also to common noise. For this purpose we consider the following modification of 
\eq{if2}:
\begin{equation}
\dot{v_i }=f(v_i )+\mu+  \sqrt{2(1-c)D}  \xi_i (t)+  \sqrt{2cD}\xi_c (t),
\label{eq:if3}
\end{equation}
where the subscript $i$ stands for the neuron's index and can attain the 
values 1 and 2. Eqs. (\ref{eq:noise1}) remain valid, and we now also have:
\be
\langle \xi_1 (t)  \xi_2 (t')\rangle=0.
\label{eq:noise2}
\ee

To characterize the correlations between the output spike trains of
two different neurons, we will use their cross-spectrum,
\begin{equation}
S_{y_1 y_2} (f) = \lim_{T\rightarrow \infty} \frac{1}{T} \langle \tilde{y}_1
\tilde{y}_2^* \rangle,  \label{eq:sy1y2}
\end{equation}
and their coherence function,
\begin{equation}
\Gamma^2 (f) = \frac{|S_{y_1 y_2 }|^2}{S_{y_1 }S_{y_2 }}.  \label{coher2}
\end{equation}

Another important measure of correlation between two spike trains is
the  correlation coefficient of the spike count. 
The spike counts  $n_1$ and $n_2$ are the numbers of spikes elicited by neurons 1 and 2, respectively,  over a time window of length $T$. Their correlation coefficient is defined as:
\be
\rho_T =\frac{\lr{n_1 n_2} -\lr{n_1 }\lr{n_2 }} {\sqrt{\lr{n_{1}^2 }-\lr{n_1 }^2}\sqrt{\lr{n_{2}^2 }-\lr{n_2 }^2}}
\ee
and its range lies between -1 and 1. In the important limit of large time
windows, one can prove the following relation between this correlation
coefficient and the zero frequency values of the cross- and power-spectra of
the spike train \cite{RocDoi07}:
\be
\rho \equiv \lim_{T \rightarrow \infty} \rho_T = \lim_{f \to 0} \frac{S_{y_1
    y_2}(f)}{\sqrt{S_{y_1}(f) S_{y_2}(f)}}.
\label{eq:rho}
\ee

\subsection{Small input correlation}

We now calculate $S_{y_1 y_2} (f)$ 
in the case of small $c$. Using \eq{lin_resp}, we obtain:
$$
S_{y_1 y_2} (f) =  \lim_{T\rightarrow \infty} \frac{1}{T}  \la \la \la
\tilde{y}_1 \tilde{y}_2^*  \ra_{\xi_1} \ra_{\xi_2}\ra_{\xi_c} = 
$$
\be
 \lim_{T\rightarrow \infty} \frac{1}{T} \la \la \tilde{y}_1 \ra_{\xi_1} \la \tilde{y}_2^* \ra_{\xi_2}\ra_{\xi_c} = 2cD
|\chi_{D,\mu}|^2,
      \label{eq:ans_sy1y2}
\end{equation}
where the averages were taken first over $\xi_1$ (with frozen $\xi_2$ and
$\xi_c$), then over $\xi_2$ (with frozen $\xi_c$), and finally over
realizations of $\xi_c$.
\eq{ans_sy1y2} has been already used in the literature
\cite{DoiLin04,LinDoi05}.

Since the limit of $\chi$ at zero frequency is given by $\frac{dr}{d\mu}$, 
for small input correlation \eq{rho} has the simple form \cite{RocDoi07}:
\be
\rho = \frac{2cD\left|\frac{dr}{d\mu}\right|^2}{rR^2}.
\label{eq:rho2}
\ee

Eqs. (\ref{coher}), (\ref{eq:syxi2}), (\ref{eq:rho}), and (\ref{eq:ans_sy1y2}) imply that, for 
small $c$, the correlation coefficient of the spike count 
is equal to the limit value of the coherence function $\gamma^2$ 
at zero frequency.

\subsection{Results}

\begin{figure}[h]
\includegraphics[width=0.45\textwidth]{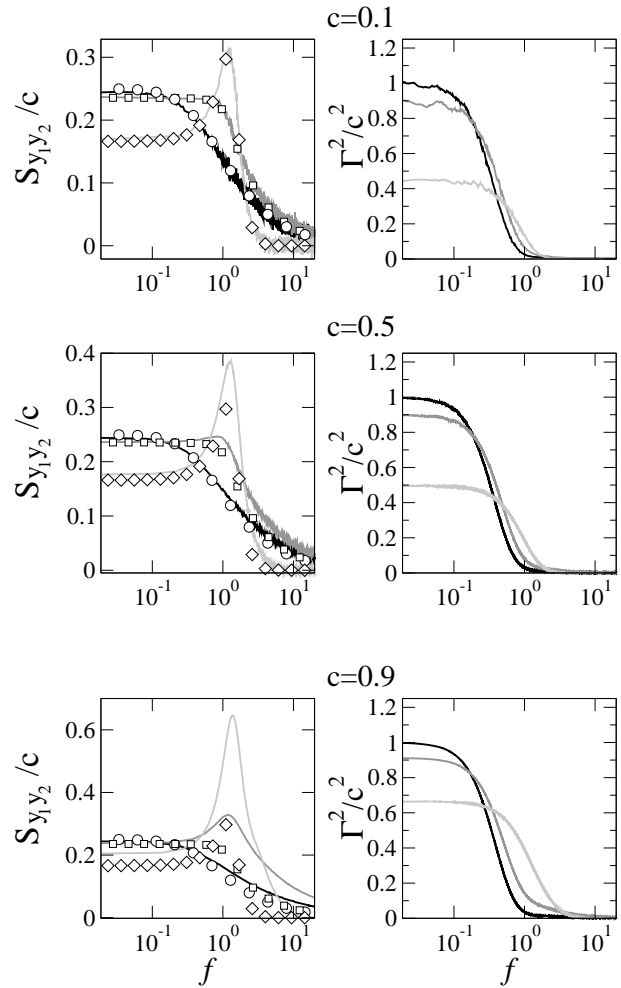}
\caption{
\label{fig:regC} 
Cross-spectra between the two output spike trains of neurons under
common noise stimulus divided by the input correlation $c$ (left
panels) and coherence function of the two output spike trains divided
by $c^2$ (right panels) in Regime C. PIF (black), LIF (dark gray), and
QIF (light gray) are compared. In the left panels, the circles (PIF),
squares (LIF), and diamonds (QIF) correspond to the
prediction from \eq{ans_sy1y2}.  }
\end{figure}

\begin{figure}[tbh]
\includegraphics[width=0.45\textwidth]{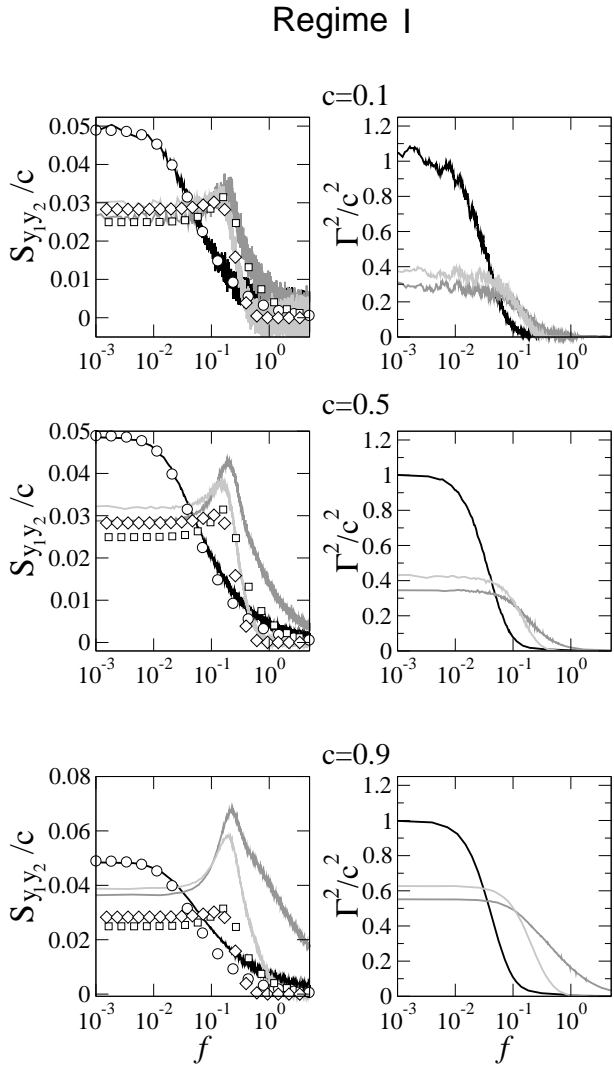}
\caption{
\label{fig:regI} 
Cross-spectra between the two output spike trains of neurons
under common noise stimulus divided by the input correlation 
$c$ (left panels) and coherence function of the two output spike trains 
divided by $c^2$ (right panels) in Regime I. PIF (black), LIF (dark gray), and QIF (light gray)
are compared. In the left panels, the circles (PIF), squares (LIF), 
and diamonds (QIF) correspond to the prediction from \eq{ans_sy1y2}.
}
\end{figure}

\begin{figure}[tbh]
\includegraphics[width=0.45\textwidth]{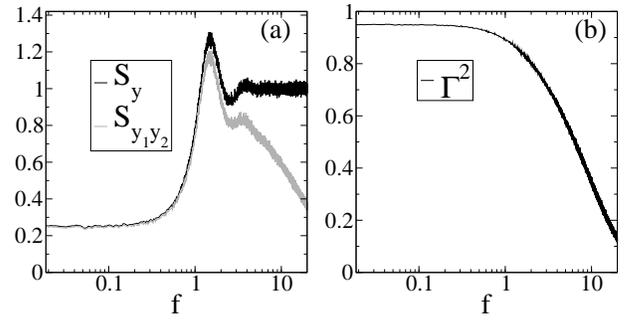}
\caption{
\label{fig:c0_999_QIF} 
Power- and cross-spectra of the output spike trains of two QIF
neurons under common noise stimulus (a) and coherence function of
the two output spike trains (b) in Regime C for $c=1-\epsilon$, with $\epsilon=10^{-3}$.}
\end{figure}

Our analysis of the two-neuron correlation relies primarily on simulations of
the stochastic differential equations (\eq{if3}) and the evaluation of the
cross-spectra (\eq{sy1y2}). This is computationally considerably more
demanding than the simple integration of the Fokker-Planck equation and the
evaluation of the analytical expressions leading to the results presented in
the previous sections. For this reason we now restrict ourselves to the
analysis of regimes C and I only. However, this suffices to lead us to three
important conclusions, which we now state.  First, as Figs. \ref{fig:regC} and
\ref{fig:regI} show, linear response theory leads to good approximations for
the cross-spectra $S_{y_1 y_2}$ for small $c$ (e.g., c=0.1). Second, we note
from \bi{c0_999_QIF} that, as the input correlation $c$ approaches 1, the
convergence of $S_{y_1 y_2 }$ to $S_y$ and of $\Gamma^2 (f)$ to the maximum
value 1 (for all $f$) is very slow. This convergence is more pronouncedly slow
for large frequencies, which corresponds to the fact that a tiny
amount of individual noise
is enough to produce a finite difference in the spiking times of the
two neurons.
Third, the coherence function in the important limit of small frequencies is
larger for the LIF in regime C as compared to the QIF and, conversely, larger
for the QIF as compared to the LIF in regime I.

In view of the discussion in \sect{2nA}, we conclude that the
correlation coefficient $\rho$ of the spike count in the limit of large time
windows is larger for the LIF than for the QIF in regime C and larger for the
QIF than for the LIF in regime I. In \bi{corr_vs_c}, this is shown to occur
for $c$ in the whole range $0\leq c \leq 1$.  We also observe that an
approximately linear dependence holds in a fairly broad range in Regimes C and
I for both models.

In order to provide a complete picture of the correlation coefficient
not only for two specific regimes but rather in a fairly broad region
of the parameter space, we show in \bi{l} and \bi{q} the ratio
$\rho/c$ as a function of $\mu$ and $D$ for the LIF and QIF,
respectively. They were estimated on the basis of the Eq.
(\ref{eq:rho2}) and are expected to be correct for small input
correlation $c$.  For the LIF, we have calculated the terms in
\eq{rho2} from the exact analytical expressions (see e.g.
\cite{VilLin09}).  For the QIF we resorted to the numerical algorithm
of Refs. \cite{Ric07} and \cite{Ric08}.

We note from that for both LIF and QIF the correlation coefficient
falls sharply when the Poissonian firing regime (low $D$ and $\mu$) is
approached. Also remarkable is the fact that, at least in the studied
parameter regions, the correlation coefficient for the LIF can
approach 1 (if $\mu$ and $D$ are large), but the correlation
coefficient for the QIF seems to have a considerably smaller upper
bound (below 0.7). Let us now describe some simple limits of $\rho/c$.
For the QIF, this quantity approaches the value $2/3$ in the limit of
strong input ($\mu>0$) and weak noise ($D\ll 1$). In the excitable
regime ($\mu<0$) at weak noise ($D\ll |\mu|^{3/2}$), i.e., when the
firing is close to Poissonian, $\rho/c$ approaches zero.

In \bi{ratio}, we show the ratio between the correlation coefficients of LIF
and QIF. The correlation coefficient is larger for the LIF in most parts of the
analyzed parameter space. Only when $\mu$ and $D$ are small, i.e., when
the firing is close to Poissonian, is the correlation coefficient larger for
the QIF than for the LIF.

Finally we turn to the simplest case of the PIF. For this model, one can show
that the correlation coefficient is given simply by $c$. In the
terminology introduced in \cite{RocDoi07}, the  correlation
susceptibility is equal to 1 for the PIF. Using \eq{ans_sy1y2}, we obtain
\be
\rho^{(\mbox{PIF})}= \lim_{f \to 0} \frac{2cD|\chi(f)|^2}{S_y (f)}.
\label{eq:rho_pif}
\ee
Using Eqs. (\ref{coher}) and (\ref{eq:syxi2}), as well as the fact that
$ \lim_{f \to 0} \gamma^2 (f)=1$ for the PIF (see \cite{SteFre72}), we obtain 
\be
\rho^{(\mbox{PIF})}=c.
\label{eq:rho_pif2}
\ee
Remarkably, this linear law, in principle valid only for small $c$, can be
shown for the PIF to be valid for all $c \in [0,1]$, as we show in 
the Appendix.

\begin{figure}[tbh]
\includegraphics[width=0.4\textwidth]{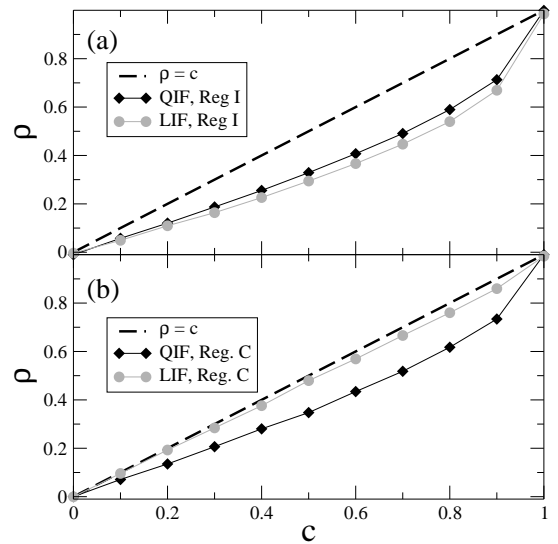}
\caption{
\label{fig:corr_vs_c} 
Correlation coefficient of the spike count vs input correlation for large time
window for LIF and QIF in firing regimes I (a) and C (b). 
}
\end{figure}

\begin{figure}[tbh]
\includegraphics[width=0.45\textwidth]{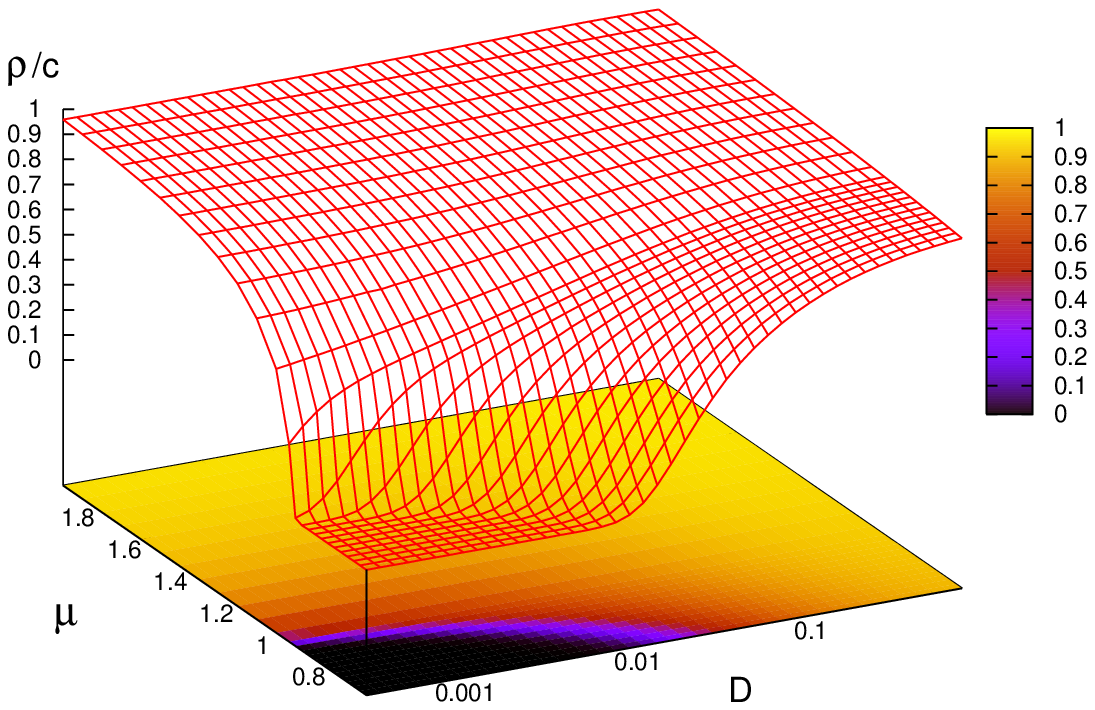}
\caption{
\label{fig:l} (Color online) 
Correlation coefficient (divided by $c$) of the spike counts at large time windows for the LIF
as a function of both $D$ and $\mu$. }
\end{figure}

\begin{figure}[h]
\includegraphics[width=0.45\textwidth]{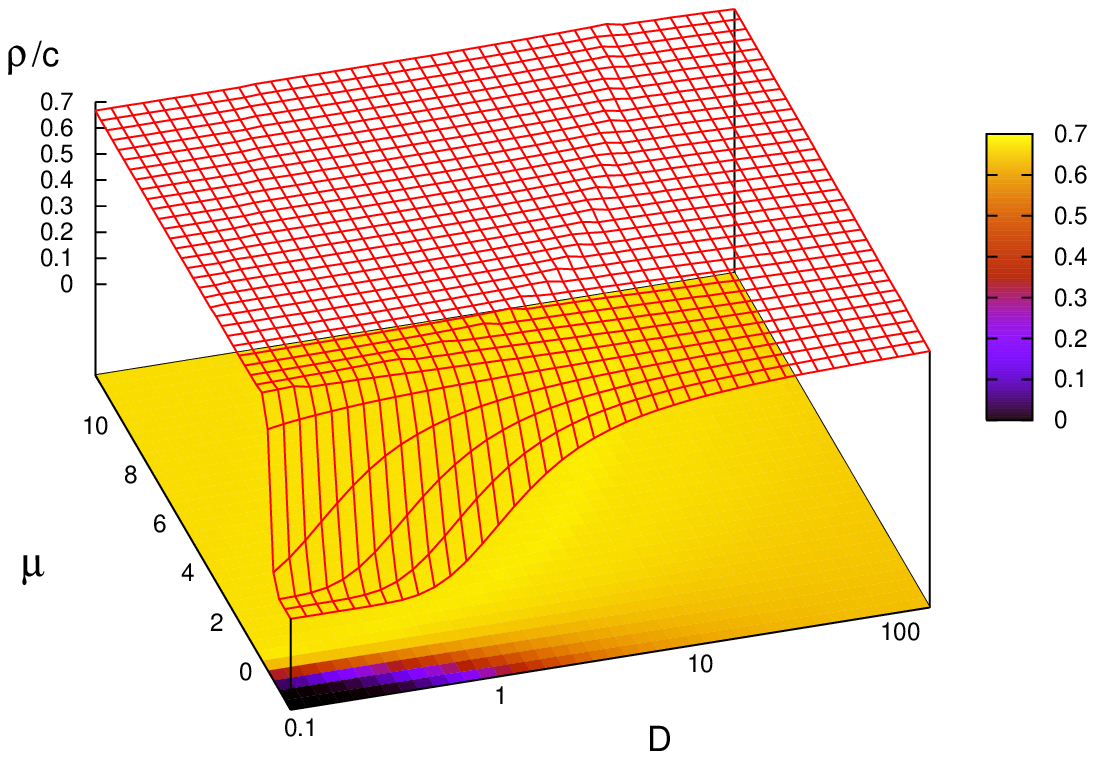}
\caption{
\label{fig:q} 
(Color online) Correlation coefficient (divided by $c$)  of the spike counts at large time windows for the QIF
as a function of both $D$ and $\mu$.  }
\end{figure}

\begin{figure}[h]
\includegraphics[width=0.45\textwidth]{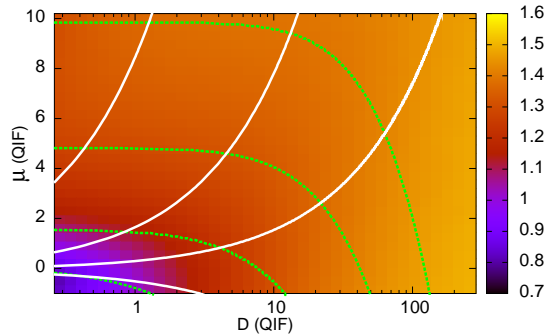}
\caption{
\label{fig:ratio} (Color online) 
Ratio between the correlation coefficients of the spike counts of LIF and QIF
in the limit of large time windows.  The contour lines for rate and CV are the
same as those depicted in Fig. 1.}
\end{figure}


\section{Conclusions} \label{sec:conc}

We have provided an extensive comparison of three important IF models in
different firing regimes, as determined by given firing rate and CV.  We have
shown that the spontaneous activity of the LIF and QIF neurons virtually
coincide in regimes characterized by weak input noise and deviates moderately
for larger values of the input noise. The dynamical response behavior,
however, strongly differs among different models, even in the same firing
regimes. This was discussed in the limit of large stimulus frequencies in
\cite{FouHan03} and extended here for the entire frequency range.

An important feature of the single neuron's response characteristics
is that, depending on the firing regime, it can be stronger at a given
frequency for the LIF as compared to the QIF or the other way around.
We have shown that this implies, as long as the linear response theory
holds true (i.e., for small $c$), that either two LIF or two QIF
neurons can display larger low-frequency correlations when driven in
part by common noise. Altogether our findings indicate that the
successful use of a certain IF model to reproduce the spontaneous
activity of biological neurons does not at all guarantee that the
correlations in the activity of a population of such neurons will be
also well described.  More important for the latter are the dynamical
response characteristics of the single neuron to an external stimulus.

We have also characterized a large portion of the physiologically relevant
parameter space of the studied IF models and concluded that the correlations
in the spike trains of two LIFs, as characterized by the correlation
coefficient for the spike count, are in most cases larger than the
corresponding correlations of two QIFs. An important exception, however,
exists: when the firing approaches the Poissonian regime, the correlations
between QIF neurons become larger than those of two LIF neurons.

It constitutes an interesting open problem to perform the same studies as done
here for other IF models, e.g. the exponential IF model introduced in
\cite{FouHan03}.  Finally, an extensive comparison between type I
\cite{RinErm89,GutErm98,LinLon03} and type II \cite{Izh01, BruHak03,
RicBru03} neurons as regards to spontaneous activity, dynamical response, and
two-neuron correlation under common stimulus is still lacking and is expected
to shed light on the problem of how large neuronal populations encode and
transmit information.
\section{Acknowledgment}
We would like to thank Magnus J. E. Richardson for help on his algorithm from \cite{Ric08}
prior to publication.

\begin{appendix}
\section{Correlation coefficient for the PIF for arbitrary $c$}
\label{app}

Here we show that the correlation coefficient for the PIF is for an
input correlation $c\in [0,1]$ given by:
\be
\rho^{(\mbox{PIF})}=c.
\label{eq:rho_pif3}
\ee
For this purpose, we track the {\em non-reset} voltage of the PIF
described by:
\be
\dot{v_i }=\mu+  \sqrt{1-c}  \xi_i (t)+  \sqrt{c}\xi_c (t),
\label{eq:eq_pif}
\ee
and observe that, in the limit of large times, the relative error in
approximating the spike count $n_i (t)$ by this unresetted voltage $v_i (t)$ approaches
zero, i.e.:
\be
\lim_{t \to \infty}\frac{n_i (t)-v_i (t)}{(n_i (t)+v_i (t))/2}=0. 
\label{eq:lim_nv}
\ee
Eq. (\ref{eq:lim_nv}) holds true because the difference between the
spike count and the non-reset voltage is a number between 0
and 1 (assuming $v_{th}- v_r = 1$). In other words, the numerator
appearing in the limit of \eq{lim_nv} remains bounded between 0 and 1
for all $t$, while the denominator goes to infinity as $t\rightarrow
\infty$.

Approximating the spike count $n_i (t)$ by this unresetted voltage
$v_i (t)$, we obtain an alternative formula for the correlation
coefficient for the spike count of the PIF:
\be
\rho^{(\mbox{PIF})} = \lim_{t \to \infty} \frac{\lr{v_1 (t) v_2 (t)} -\lr{v_1 (t)}\lr{v_2 (t)}}{\sqrt{(\lr{v_{1}^2 (t)} -\lr{v_1 (t)}^2)(\lr{v_{2}^2 (t)} -\lr{v_2 (t)}^2)}}.
\label{eq:rho_v}
\ee
To calculate the right-hand side of \eq{rho_v}, we use the formal solution of
\eq{eq_pif}, given by:
\be
v_i (t)=\int_{0}^t (\mu  + \sqrt{1-c}  \xi_i (t')+  \sqrt{c}\xi_c (t'))dt'.
\label{eq:v_pif}
\ee
Substituting \eq{v_pif} in \eq{rho_v} and using Eqs. (\ref{eq:noise1}) and
(\ref{eq:noise2}), we obtain \eq{rho_pif3}.

\section{Numerical methods}
\label{app2}

The contour lines for the rate for both LIF and QIF, as well as the
respective power spectra, were obtained by resorting to the numerical
algorithm of Ref. \cite{Ric08}.  For the LIF (QIF), the rate and CV
were calculated for the points on a roughly $2000\times 2000$ ($10^3
\times 10^3$) grid over the $(D,\mu)$ region displayed in Fig.  1.
The integration step for the numerical evaluation of the Fokker-Planck
equation (see \cite{Ric08}) was equal to $10^{-4}$ for the LIF and
$10^{-3}$ for the QIF, with the threshold and reset at $\pm \infty$
being numerically replaced by $\pm 500$ for the latter.  The CV was
estimated by assuming that the power spectrum at a frequency $10^3$
times smaller than the firing rate was equal to $r R^2$
(see \eq{lim_f_0}) for both LIF and QIF.  

The gain and phase for the linear response of LIF and QIF were
computed using the algorithm of Ref. \cite{Ric07}, with the same
integration steps (as well as reset and threshold values for the QIF)
as above. Finally, the cross-spectra $S_{y_1 y_2}$ were obtained from
a fast Fourier transform algorithm after integrating the stochastic
diferential equations \eq{if3} with a time step of $10^{-3}$. For the LIF, a correction based on the probability that
the voltage reached the threshold and decreased below it within the
time interval $dt$ was implemented (see \cite{LinLon05}).

\end {appendix}

 \end{document}